\documentclass[journal]{vgtc}         % review (journal style)

\ifpdf%                                % if we use pdflatex
  \pdfoutput=1\relax                   % create PDFs from pdfLaTeX
  \usepackage{graphicx}                % allow us to embed graphics files
  \DeclareGraphicsExtensions{.pdf,.png,.jpg,.jpeg} % for pdflatex we expect .pdf, .png, or .jpg files
\else%                                 % else we use pure latex
  \usepackage{graphicx}                % allow us to embed graphics files
  \DeclareGraphicsExtensions{.eps}     % for pure latex we expect eps files
\fi%

\graphicspath{{figures/}{pictures/}{images/}{./}} % where to search for the images

\usepackage{microtype}                 % use micro-typography (slightly more compact, better to read)
\PassOptionsToPackage{warn}{textcomp}  % to address font issues with \textrightarrow
\usepackage{textcomp}                  % use better special symbols
\usepackage{mathptmx}                  % use matching math font
\usepackage{times}                     % we use Times as the main font
\usepackage{cite}                      % needed to automatically sort the references
\usepackage{tabu}                      % only used for the table example
\usepackage{booktabs}                  % only used for the table example
\usepackage{subfigure}                 % subfigures
\usepackage{comment}
\usepackage[linesnumbered,ruled]{algorithm2e}
\usepackage{hyperref}
\usepackage{enumitem}
\usepackage{multirow}

\DeclareMathAlphabet{\mathcal}{OMS}{cmsy}{m}{n}

\usepackage{todonotes}                      % ToDo Notes
\usepackage{tikz}
\usetikzlibrary{arrows,arrows.meta,patterns,calc,shapes,decorations.pathreplacing}
\usepackage{adjustbox}
\usepackage{balance}

\vgtccategory{Research}
\vgtcpapertype{application/design study}

\title{Designing Feature Vector Representations:\\ A case study from Chemistry}

\author{Signe Sidwall Thygesen$^\dag$, Daniel Witschard$^\dag$, Andreas Kerren, Talha Bin Masood, and Ingrid Hotz}
\authorfooter{
\item
 S.S. Thygesen, A. Kerren, Talha Bin Masood and Ingrid Hotz are with Link\"oping University, Sweden. Email: \{signe.sidwall.thygesen\,$|$\,andreas.kerren\,$|$\,talha.bin.masood\,$|$\,ingrid.hotz\}\break@liu.se.
\item
 D. Witschard is with the Linnaeus University, V\"axj\"o, Sweden. E-mail: daniel.witschard@lnu.se.\break
\item[$^\dag$]S.S. Thygesen and D. Witschard contributed equally to this work.
}

\abstract{We present a case study investigating feature descriptors in the context of the analysis of chemical multivariate ensemble data. 
The data of each ensemble member consists of three parts: the design parameters for each ensemble member, field data resulting from the numerical simulations, and physical properties of the molecules. 
Since feature-based methods have the potential to reduce the data complexity and facilitate comparison and clustering, we are focusing on such methods.
However, there are many options to design the feature vector representation and there is no obvious preference.
To get a better understanding of the different representations, we analyze their similarities and differences. 
Thereby, we focus on three characteristics derived from the representations: the distribution of pairwise distances, the clustering tendency, and the rank-order of the pairwise distances.
The results of our investigations partially confirmed expected behavior, but also provided some surprising observations that can be used for the future development of feature representations in the chemical domain.
}

\vgtcinsertpkg

\begin{document}
\maketitle

%--------------------------------------------------------------
\section{Introduction}
%--------------------------------------------------------------

Feature-based data analysis is a commonly used approach in scientific visualization, for example to reduce data complexity or to make the data accessible for statistical analysis~\cite{Hotz2020, Whitaker2020}. Features can initially be anything that presents interesting information from a dataset in a compressed form. Sometimes features are directly motivated by domain-specific questions, like vortices in the flow analysis~\cite{Post2003}, however, this is not the case and more generic feature definitions are used, e.g., using topological representations~\cite{yan2021scalar}, or abstract descriptors resulting from an automatic feature generator~\cite{Katz2016}.
Generally, there is not a single correct choice, but many plausible alternatives.
The specific choice of a feature criterion is an important step that can have a significant impact on the analysis outcomes and consequently on a visualization system.

In this paper, we use a case study to illustrate how one can utilize feature vector comparison approaches to support this step.
Thereby, we were driven by questions like: \emph{What is a suitable way to quantitatively describe a feature?} \emph{What constitutes a good feature vector representation, and how can they be compared?} \emph{Do they provide the same view on the data or do they tell different stories?}, and finally, \emph{How important is the domain knowledge in this process?}
The work has been inspired by a recent visualization project in a chemistry application, specifically a level-of-detail exploration of electronic transition ensembles using hierarchical clustering~\cite{Thygesen2022}. 
The visualization system has been designed in a participatory design process and the result has been considered successful by all partners.
The system targets ensemble data, data that consists of many similar datasets (referred to as ensemble members) specified by a set of parameters, that shall be compared and jointly analyzed.
A central element in the analysis pipeline is a feature vector that represents one ensemble member and builds the basis for hierarchical clustering and level of detail visualizations in the system.  The feature vector definition was guided by the attempt to stay as close as possible to the commonly used analysis methods of the chemists. 
While the system is now regularly used in the work of our collaborating chemists, we observed that not all parts are used equally. Especially, one of our original incentives, the prediction of physical properties from the simulation data, does not seem to be as prominent as we were hoping. There could be different reasons for this, it could be that there is no strong correlation between the charge transitions and the physical properties, or, what is more likely, the chosen feature vector is not optimal for this purpose.
This motivated us to consider and compare other alternative feature vectors.
Thereby, we tried to be open and consider feature vectors with very different characteristics and dimensionality also neglecting the domain specific advice and common practices from the chemists. The results, which were partially expected but partially also surprising, are reported in this paper.

Similar discussions can frequently be found in the field of visual analytics (VA) \cite{VA1,KS:12} and several approaches and tools have been proposed to explore and compare feature vector representations, there typically referred to as embeddings. 
One motivation of these works is to get a better understanding of the properties of embedding spaces. For example, Smikov et al.~\cite{EMBCOMPARE1} are interested in embeddings in machine learning applications, whereas Liu et al.~\cite{EMBCOMPARE2} propose methods for visual exploration of semantic relationships in neural word embeddings.
Other techniques focus on the comparison of different embeddings, which include ``Parallel Embeddings'', a generalization of classical parallel coordinates to sequences of  representations~\cite{EMBCOMPARE4}, or an approach comparing embeddings by capturing similarity between word and document embeddings~\cite{EMBCOMPARE3}.
In our case study, we make use of the interactive visual analytics prototype implemented by Witschard et al. \cite{witschard2022}. 
%In our case study, we make use of some of these methods, implemented in the interactive visual analytics prototype by Witschard et al. \cite{witschard2022}. 
We are especially interested in non-parametric comparison methods~\cite{Wasserman2006} that can abstract from the different metric spaces as, e.g., Spearman's rank-order correlation methods based on sorted pairwise distances~\cite{Fieller1957}.
In general, these methods are not targeted toward the typical analysis process in a scientific domain like chemistry, where we do not have a ground truth and the data analysis is an essential part of the knowledge discovery process. 
But even though the prototype is more targeted towards addressing the case of text similarity, it turned out to be very useful in the context of scientific visualization.
%However, it turned out they have also can be very useful in that context.

With this case study, we do not target the domain scientists from the application (here the chemists) as users, but mainly visualization researchers designing feature-based visualization systems like us.
While the case study illustrates the impact of the definition of a feature vector in one specific application, similar questions arise in many other scientific visualization applications as well.

%--------------------------------------------------------------
\vspace{1mm}
\section{The Application Case}\label{sec:background}
%--------------------------------------------------------------

\begin{figure}[b]
    \centering
    \includegraphics[width=1.0\linewidth,trim=80 10 110 90, clip]{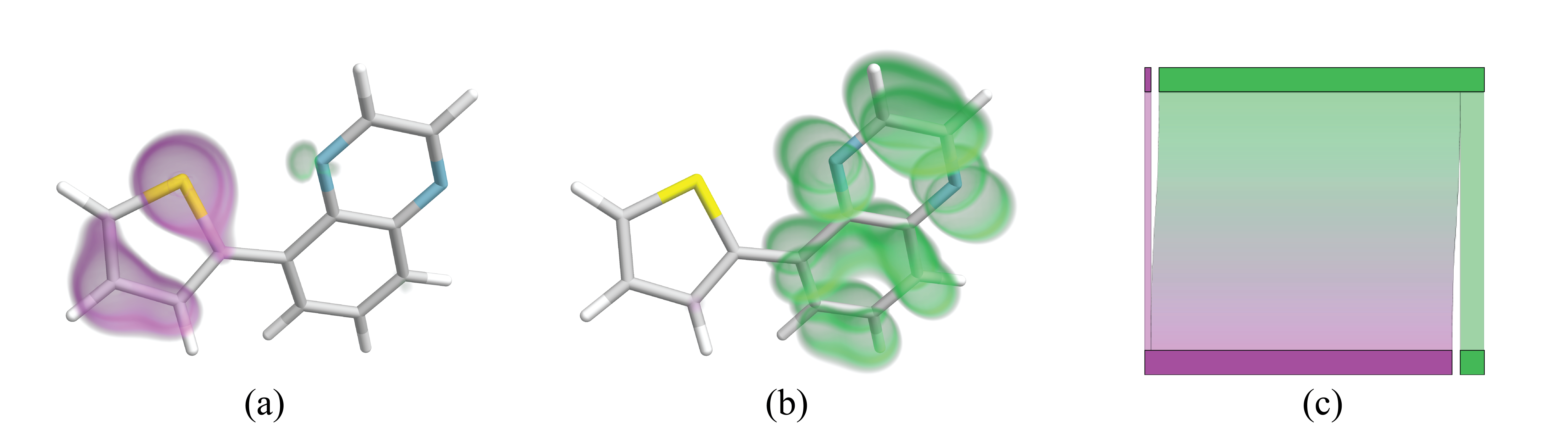}
%    \includegraphics[width=0.9\linewidth]{Figures/mol_rendering.png}
%    \includegraphics[width=0.34\linewidth]{Figures/au-phe-phephe-td_State1-holeAndParticle_isosurfaces-01}
%    \hspace{0.08\linewidth}
%        \includegraphics[width=0.34\linewidth]{Figures/au-phe-phephe-td_State1-holeAndParticle_isosurfaces-02}
    \caption{An electronic transition is described by a pair of scalar fields ((a) before and (b) after transition), 
    usually visualized as two isosurface images with the molecular context for a side-by-side comparison. The colors represent two molecular subgroups.
    (c) The corresponding transition diagram directly visualizing the Charge Transfer Matrix, one of the feature vectors used to describe the transition.}
    \label{fig:electronic-transition-isosurfaces}
\end{figure}

In our application case, chemists are interested in changes in the electronic structures of molecules when interacting with light, called an \emph{electronic transition}. 
Studying such transitions is crucial in the context of designing novel materials with desired physical properties.
For this purpose, chemists perform numerical simulations for large sets of different molecules constituting our ensemble.  The resulting data for each molecule, one ensemble member, consists of three main parts:
\begin{itemize}
\item[(i)] The model parameters for each ensemble member, in our case the molecular configuration.
\item[(ii)] The simulation results, often multi-fields, in our case describing the electronic transition given by a pair of scalar fields capturing spatial electron density distribution before and after the transition~\cite{Martin2003NTO, eurovis_electronic_densities}. 
\item[(iii)] Predicted properties, in our case physical properties associated with the transition, relevant for the practical relevance of the materials.
\end{itemize}
This is a structure that can be found in many applications. Often the goal of such simulations is to understand the relation between these three parts of the data.
In our case, the chemists are interested in understanding any relationships between the physical properties, the model parameters, and the changes in electronic structure during a transition. 
Usually, for each ensemble member, the pair of fields is analyzed by comparing isosurface visualizations of the two fields~\cite{AlSaadon2019,Stone2011}, as shown in Figure~\ref{fig:electronic-transition-isosurfaces}(a, b). 
In this way, an efficient comparison of large ensembles is hardly possible~\cite{Haranczyk2008}.
To enable better ways of comparison and exploration, a strategy is to create quantitative feature vector representations for each ensemble member, which is the focus of this paper.

%--------------------------------------------------------------
\section{The Feature Representations}\label{sec:feature}
%--------------------------------------------------------------

For this case study, we have selected a set of different feature vector representations summarized in Table~\ref{table:representations} that capture different aspects of the data, mainly focusing on the simulation results and the physical properties. 
The first two representations closely reflect typical analysis approaches used by the chemists. The next two representations ignore the common domain specific approaches, but still represent the simulation result. The last two vectors are simple encodings of the physical properties.
To encode model parameters specifying the individual ensemble members, we decided not to use a feature vector but apply a labeling strategy, which is summarized in Table~\ref{table:labels}.

%----------------------------------------------------------------
\paragraph*{Transfer-based quantification using molecular subgroups.}
%defines a feature vector that is directly inspired by the analysis methods used by the Chemists.
Using molecular subgroups to analyze the two scalar fields of the transition is a common approach in chemistry to reduce the complexity of the data. Typically, the number $M$ of groups lies between two and ten.
The charge transfer is then described by the change of the total charge in a few spatial segments, defined by the molecular subgroups, which is derived from the two scalar fields.
This change can be represented in a compact form as a $M\times M$ matrix, called the \emph{Charge Transfer Matrix (CTM)}~\cite{eurovis_electronic_densities}.
It can also be directly visualized in Charge Transfer Diagram, compare \autoref{fig:electronic-transition-isosurfaces}(c) for an example with $M=2$.
We also make use of a simplified version of this matrix representation, called the \emph{Transfer Feature Vector (TranFV)} which has been used in our original work~\cite{Thygesen2022}. The dimensionality of this representation is equal to $2M$.
%, in our case $M=3$.
%We use the Euclidean metric in these feature vector spaces.

%----------------------------------------------------------------
\paragraph*{Topological signature of the change in the scalar field.} 
Topological methods are well known to provide expressive summaries of scalar fields~\cite{Chazal2021}. 
Thereby, one studies the lifetime of topological features of isosurfaces,  when changing the isovalue. Features that are considered include components, loops, and voids of the isosurface.
We utilize such a summary to define a novel feature vector that quantifies the change in the topological structure of the two scalar fields.
A typical metric in this space is the bottleneck distance \cite{edelsbrunner2008persistent} used to quantify the change.
We refer the reader to Yan et al.~\cite{yan2021scalar} for a more in-depth explanation of topological descriptors for scalar field comparison. The resulting feature vector is three-dimensional, each component quantifying differences for one dimension of the topological invariant. 
This approach does not consider the spatial segmentation usually used by the chemists, and thus ignores domain knowledge generally used in the analysis of the data.

\begin{table}
\begin{center}
\small
\begin{tabular}{ ll p{16em}  l }
\toprule
 \textbf{Rep.} & \textbf{Data} &\textbf{Description} & \textbf{Dim.} \\ 
 \midrule
  TranFV & (ii) & Transition Feature Vector, compressed version of CTM. & $6$ ($2M$) \\  
 %\hline
 CTM & (ii) &  Charge Transfer Matrix. & $9$ ($M^2$) \\  
 %\hline
 TopoFV  & (ii)&  Topological Feature Vector, using topological difference between the fields. & $3$ \\%\talha{I believe TopoFV is not dependent on $M$} \\
 %\hline
 SFV & (ii) &  Scatterplot Feature Vector. & $121$ $(2K + 1)^2$  \\
 %\hline
 PFV & (iii) &  Properties Feature Vector, using three chemical properties (wavelength, rotatory strength and oscillatory strength). &  $3$  \\
 %\hline
 WFV & (iii) &  Wavelength Feature Vector. &  $1$  \\
 \bottomrule
\end{tabular}
\end{center}
\caption{Feature vector representations of electronic transitions used in this paper. For CTM and TranFV, the dimensionality depends on the number of subgroups in the molecules $M$, here $M=3$. The dimension of SFV depends on the number of bins, here $K=5$. 
}
\label{table:representations}
\end{table}
    
\begin{figure}[b]
    \centering
    \includegraphics[width=1\linewidth]{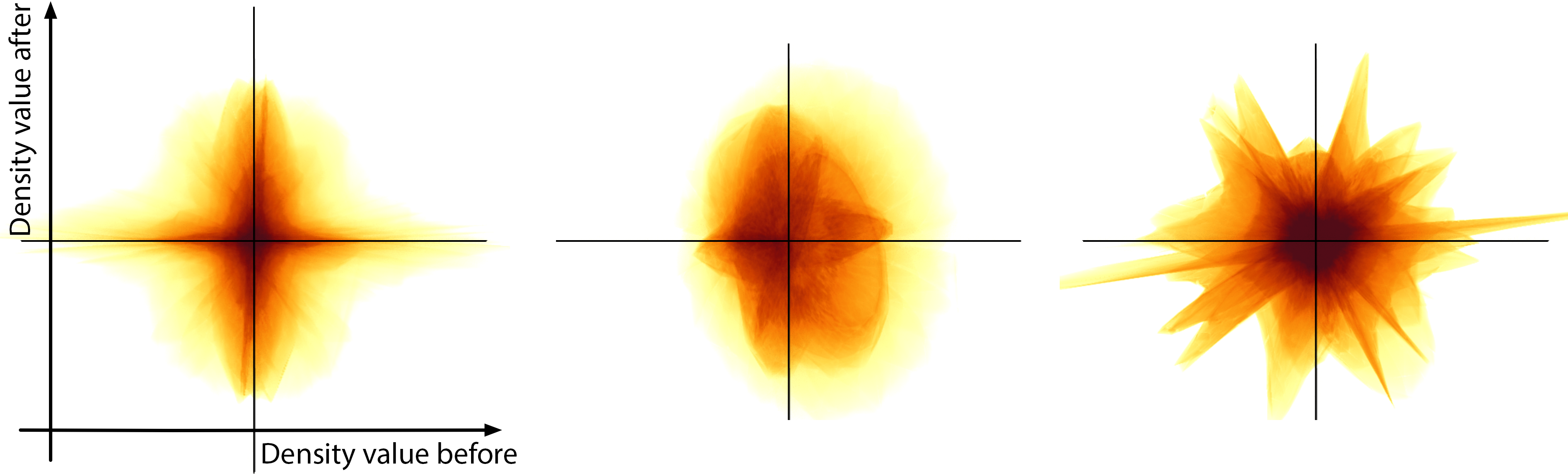}
    \caption{For a bi-variate analysis of the transition, the two scalar fields are represented in a continuous scatterplot with the charge density values before and after the transition on the axis. The image shows examples how such scatterplots can look like for three different molecular conformations. To define the feature vector, we have used a coarse representation of these scatterplots.
    }
    \label{fig:bivariate}
\end{figure}

\paragraph*{Bivariate analysis feature vector.} 
This representation uses the fact that each ensemble member in our data is a bivariate field (two scalar fields). To generate a feature vector, the values of these two fields are discretized into bins in a scatter plot, similar to the continuous scatter plot used in~\cite{Sharma2021CSPpeeling}.
\autoref{fig:bivariate} shows examples of a bivariate representation for three different molecules. 
We call this representation the \emph{Scatterplot Feature Vector (SFV)}.
The dimensionality is dependent on the number of bins $K$; we use $K=5$. We also tried $K=20$, but later disregarded this one because of its high dimensionality.

\paragraph*{Using physical properties.}
For each member in our data set, we have several physical properties describing some relevant chemical aspect \cite{norman2011perspective}.
Using these properties as a feature vector is a straightforward way of creating a quantitative representation. We call this \emph{Properties Feature Vector (PFV)}. In addition, we also try a feature vector representation only consisting of one of these properties, to test a more extreme case where the representation is one-dimensional. This is the \emph{Wavelength Feature Vector (WFV)}.

%-------------------------------------------------------------------------
\section{Comparative Measure for Feature Representations}
\label{sec:method}
%-------------------------------------------------------------------------

As described in \autoref{sec:background}, we use a set of feature vector representations for comparison and analysis. They are all quantified from an ensemble data set consisting of 180 members. The main goal of this case study is to explore the similarities and differences between these representations.
Since the dimensionality of the different feature vector representations varies a lot, we mainly focus on the comparison of orderings and groupings of ensemble members using the different representations. 
Therefore, we have chosen to include the following three characteristics in the analysis: (1) the distribution of pairwise distances, (2) the clustering tendency, and (3) the rank-order of the pairwise distances. Since we have no clear a priori knowledge of what to expect. We use available methods and tools from Visual Analytics. Furthermore, to account for differences in dimensionality and distance magnitude, we are using normalized Euclidean distances as the base for all our calculations. 
In addition to the direct comparison of orderings, we use labeling in our analysis to explore the correlation between clustering tendencies.  In our visualizations, the labels are represented as color. The labels that we use span over all parts of the data set, they are summarized in \autoref{table:labels}.

%-------------------------------------------------------------------------
\subsection{Distance distribution analysis}
A straightforward way to gain initial insights into the similarity of the point distribution in two multidimensional spaces is to check the distribution of the pairwise distances of the point pairs. For instance, if the distance distribution in space A is uni-modal and the distance distribution in space B is bi-modal, we can conclude that the representations have different characteristics. However, we must also be aware that similar distribution patterns do not necessarily imply that the two representations are similar. 

%-------------------------------------------------------------------------
\subsection{Clustering tendency}
\label{sec:clustering-tendency}
We use dimensionality reduction (DR)~\cite{Espadoto:21} with t-SNE~\cite{TSNE} to explore the clustering tendency of the feature vector sets. 
Although t-SNE is not a clustering algorithm, it still gives insights into the distribution of the data in terms of groupings.
If there are clusters in a multidimensional data set, this will likely be reflected in the t-SNE plot. 
An example of this can be seen in \autoref{fig:transition-and-charge-with-cluster} where there is a high level of consistency between the plots and the cluster identity labels from hierarchical clustering on the same data set. 
To link the feature vector representations back to properties interesting for the domain scientists, we create multiple labels based on such properties.

\begin{table}[t]
\begin{center}
\small
\begin{tabular}{ c  p{22em} }
\toprule
 \textbf{Labeling} & \textbf{Description} \\ 
 \midrule
 Metal & Metal atom in the molecule: Ag~/~Au~/~Cu. \\  
 %\hline
 Subgroup & A specific subgroup of the molecule: 6 different types. \\  
 %\hline
 PP1 & Wavelength, divided into: low, medium, high. \\
 %\hline
 PP2 & Members having the highest oscillatory strength. \\
 %\hline
 PP3 & Members having non-zero rotatory strength. \\
 %\hline
 Clusters & The set divided into 3 clusters (using agglomerative clustering based on the TranFV representation). \\
 \bottomrule
\end{tabular}
\end{center}
\caption{The labelings used in our analysis, to explore correlation with clustering tendencies. PP1-PP3 are based on physical properties (same as we have used in the PFV and WFV representations).
}
\label{table:labels}
\end{table}

%-------------------------------------------------------------------------
\subsection{Rank-order correlation}
Another characteristic obtained from sorting the pairwise distances from smallest to highest are the rank-order numbers.
We create a visual representation of the rank order correlation by plotting the rank-number from one representation on the x-axis and the rank-number from the other representation on the y-axis. 
Then, two representations are considered as consistent or correlated to each other if they have the same ``opinion'' on the order of point pairs. 
In this way, a strong correlation between the representations will be reflected as a tendency of the point cloud to approach the line $y = x$.
To quantify this relation, one can calculate the Pearson correlation coefficient, measuring the linear correlation between two variables. Given two random variables $X$ and $Y$ it is computed as $cov(X,Y)/\sigma_x\sigma_y$, where $cov$ is the covariance and $\sigma$ is the standard deviation of the respective variables.
As an alternative method, we calculate the Jaccard-index~\cite{JACCARD} measuring the similarity of two sets $A$ and $B$ as $J(A,B)=|A\cap B|/|A\cup B|$. Therefore, we consider the set of 1,000 closest points in each representation.

%-------------------------------------------------------------------------
\section{Case Study}
%-------------------------------------------------------------------------

\begin{comment}
{\color{blue} 
\begin{itemize}
    \item Scatter plots with labels
    \item Clustering tendencies: observed for TranFV, CTM and TopoFV
    \item correlation plots
    \item Most interesting results: "bottleneck" vector correlation with wavelength and metal? (And possibly oscillatory strength?)
\end{itemize}
}
\end{comment}

In this case study, we investigate the similarities of the different feature vector representations for a chemical application that have been presented in \autoref{sec:background}. Therefore, we apply the analysis methods introduced in \autoref{sec:method}. 
Thereby, we compare the different feature vectors directly to each other but also consider correlation to labels that are derived from the model parameters and physical properties. The goal of the study is that get a better understanding of the potential of the different feature vectors for the use in a visual analysis system. As such, it does not target the chemists but the designers of the visualization system.

\begin{figure}[t]
    \centering
    \includegraphics[width=\linewidth]{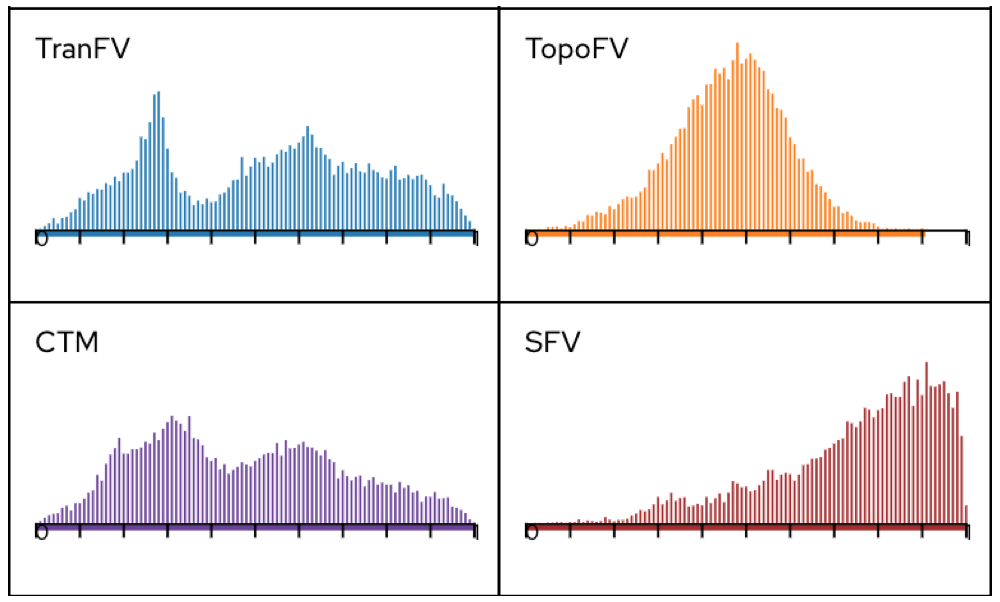}
    \caption{Visualizing the normalized distance distribution can reveal important insights to the similarity of different representations. This example shows that the distances have a similar distribution for TranFV and CTM, but that the other representations show a considerably different pattern.
    }
    \label{fig:distance-distributions}
\end{figure}

%-------------------------------------------------------------------------
\subsection{Comparing distance distributions}
At first, we want to get an overview of the distribution of the ensemble member representations in the respective feature vector spaces. In our case, every ensemble member relates to one molecular configuration. 
Their embedding/representations in the feature space is in the following referred to as points.
If the distance between a pair of points is small, the underlying ensemble members are considered similar in this representation. 
%Analyzing the distribution of the normalized pairwise distances of the points reflects properties of the embedding, e.g., if many/few points lie close/far away from each other. 
Looking at the resulting distributions, shown in \autoref{fig:distance-distributions}, we can observe different characteristics. 
In contrast to the other representations, TranFV and CTM are similar in that they both show two peaks in their distribution, which confirms their relationship as TranFV is derived from the CTM.
This could be a hint that they have a stronger clustering tendency than the other representations.
A peak at small distances can refer to point pairs lying in one cluster. The second peak at higher values would then result from distances between points belonging to different clusters.
The other distributions have a different appearance, suggesting they are not capturing the same features in their representations. Especially the distribution of SPV with mostly high values hints at a very scattered distribution of the points.

\begin{figure}[b]
    \centering
    \includegraphics[width=\linewidth]{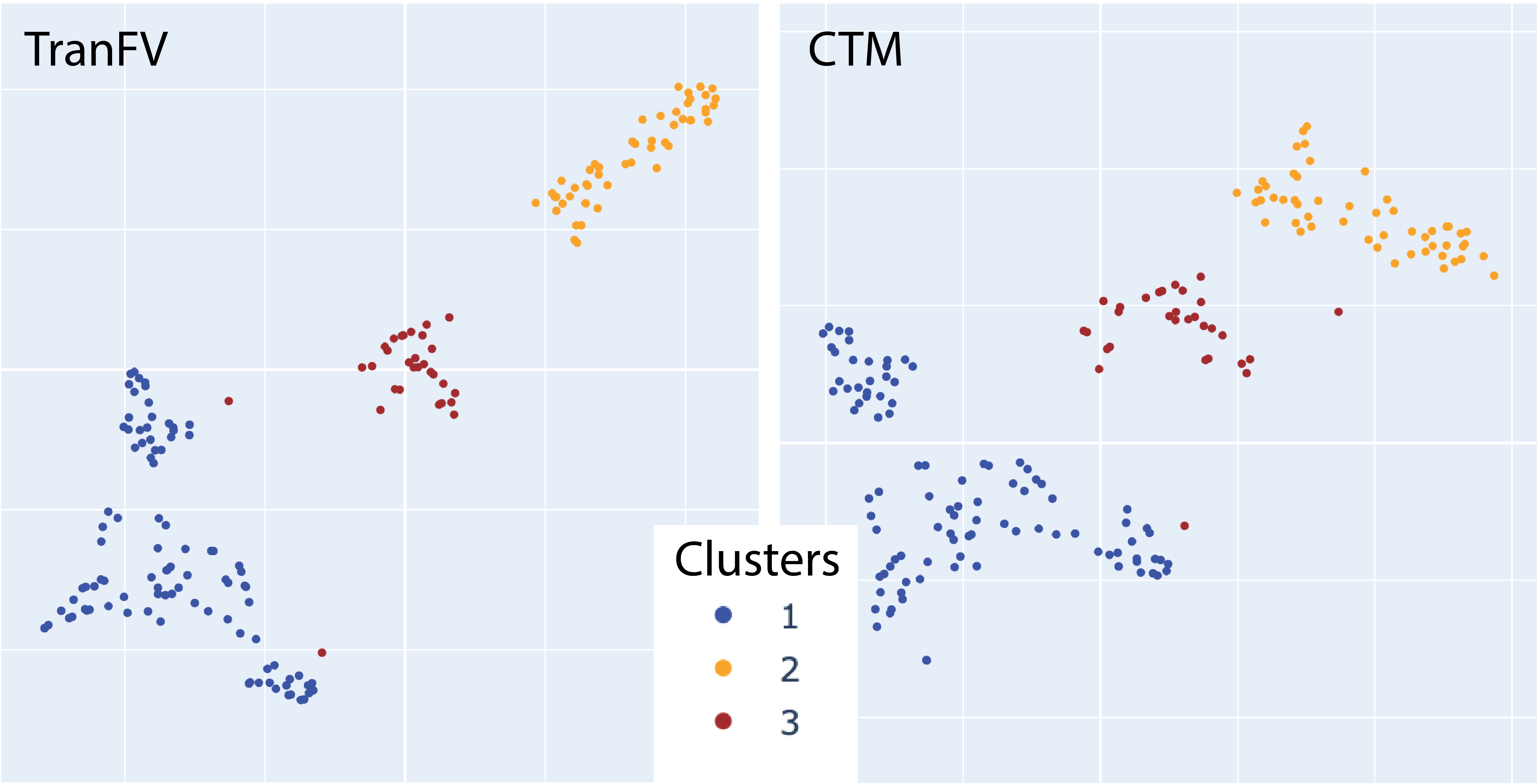}
    \caption{t-SNE plot of the TranFV and CTM representations together with the cluster labeling showing a clear correlation between the cluster tendency and the labels, for both representations.
    }
    \label{fig:transition-and-charge-with-cluster}
\end{figure}

\subsection{Investigating cluster tendencies and label correlation}
\begin{figure}[h]
    \centering
    \includegraphics[width=\linewidth]{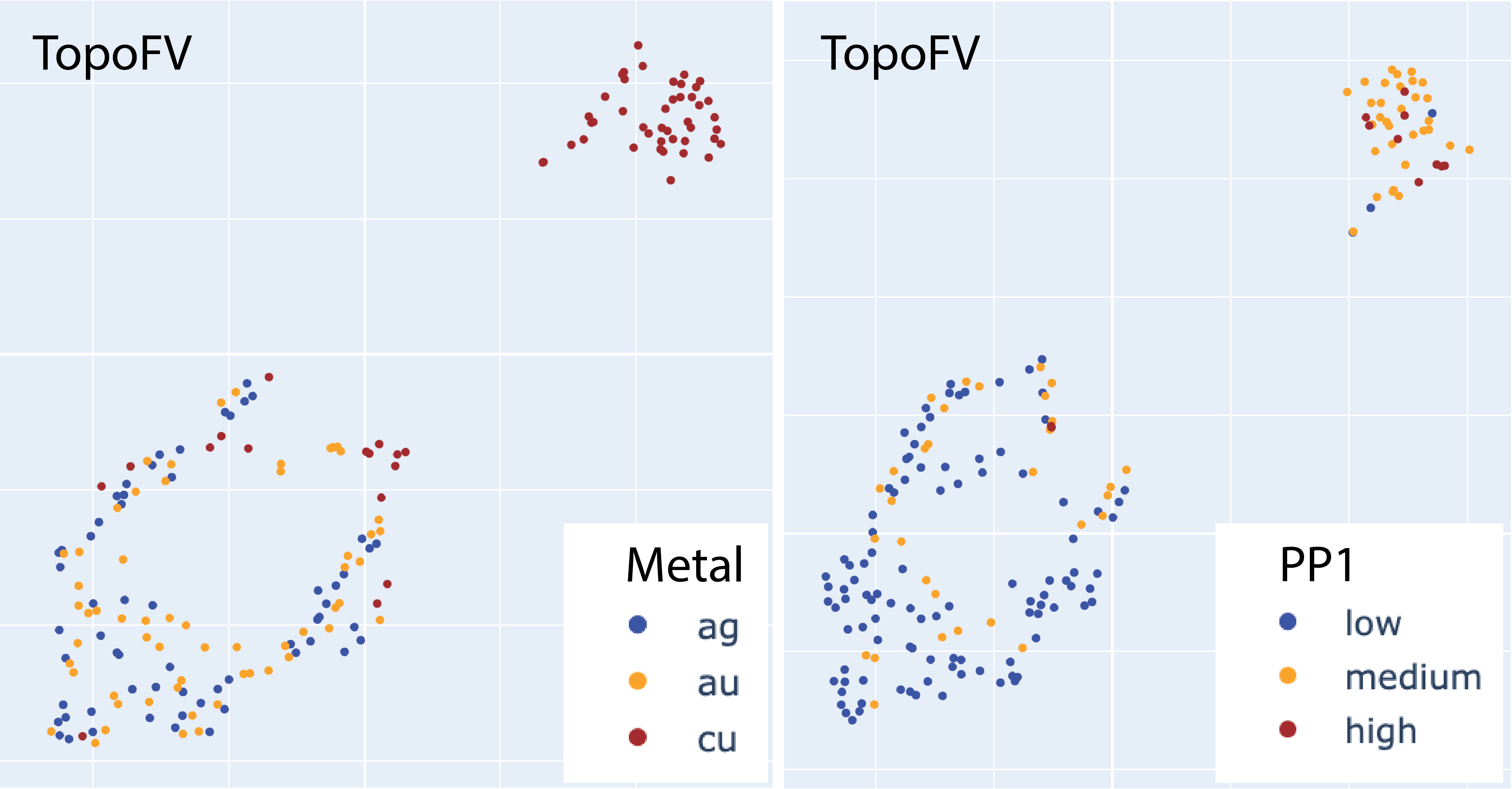}
    \caption{t-SNE plot of the TopoFV representation together with the metal labeling (left) and the PP1 labeling (right), showing there seem to be some correlation between the clusters and these labels.
    }
    \label{fig:topo-with-metal-and-wavelength}
\end{figure}

To dive deeper into exploring the groupings of the different points, we inspect their two-dimensional projection using t-SNE.  
For better understanding of the correlation of the groups resulting from different representations or to other physical properties, we apply a label-specific coloring, compare \autoref{table:labels}.
We have explored all combinations and plot our most interesting observations in \autoref{fig:transition-and-charge-with-cluster} and \autoref{fig:topo-with-metal-and-wavelength}. 

\autoref{fig:transition-and-charge-with-cluster} compares the cluster labels with the t-SNE plots of both the TranFV and CTM representations. The strong agreement for the TranFV representation is not surprising, since the labels are based on it. However, looking at the plot for CTM, we observe that this is very similar, confirming that TranFV captures similar features as CTM from the underlying data.
We observe some outliers having cluster label 3, two in each plot. Because we are using t-SNE, the distance between clusters should be interpreted with great care. However, these outliers could be interesting to investigate further.

In \autoref{fig:topo-with-metal-and-wavelength}, we observe the t-SNE plot of the TopoFV representation, labelled both by Metal and by PP1.
This reveals an interesting correlation: for the Metal label there is a separate cluster only having label `cu'; and for PP1, we see that the same cluster is having mainly labels medium and high. Combining this information gives insight into both the cluster characteristics of the TopoFV representation and the correlation with these physical properties. This suggests that this representation, built on topological invariants, is able to capture some chemically relevant features of the data even though it does not make use of chemical domain knowledge.

\subsection{Exploring correlation between representations}
\begin{figure}[h]
    \centering
    \includegraphics[width=\linewidth]{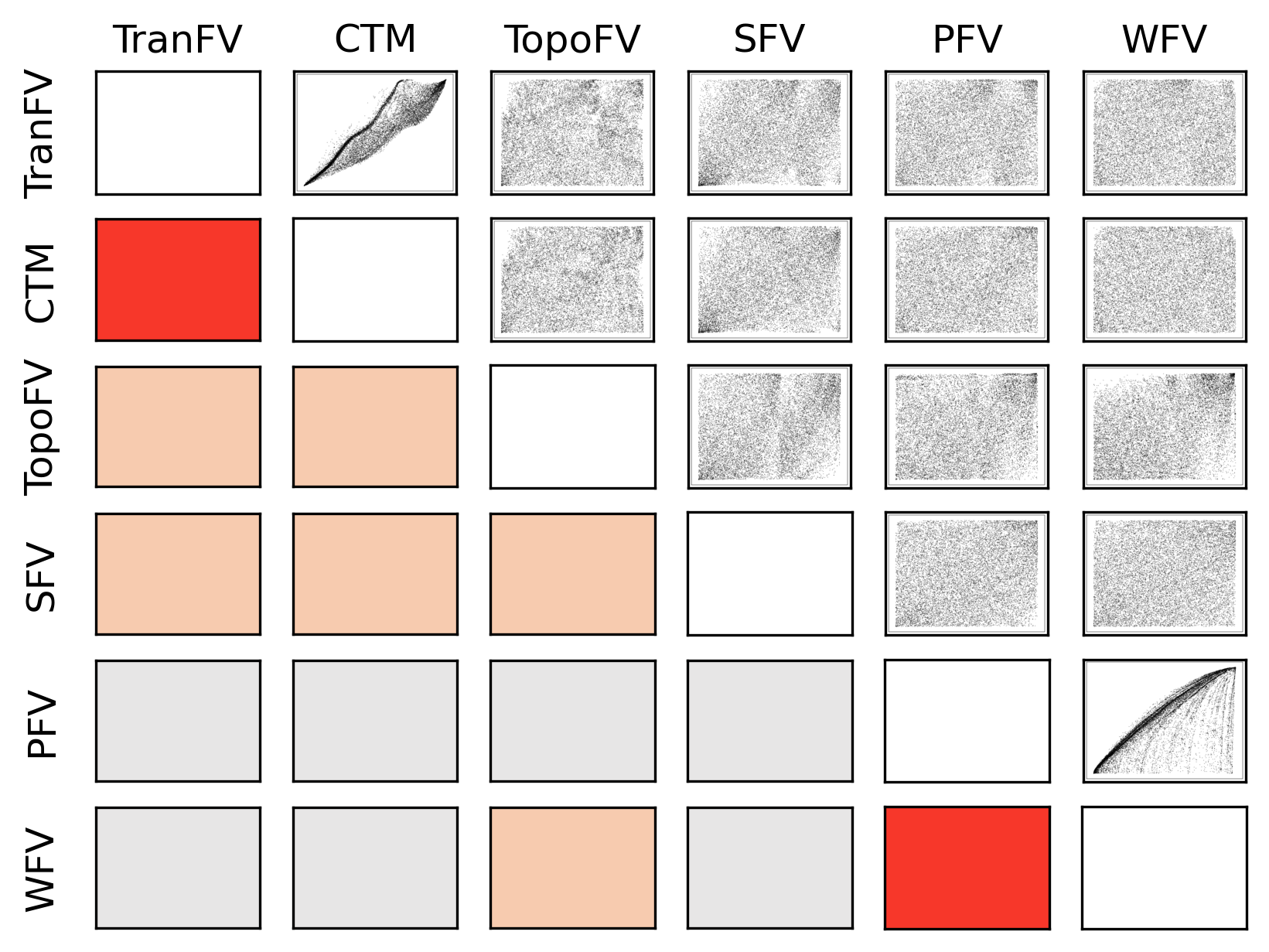}
    \caption{A scatterplot matrix showing the correlation between the representations. The upper half shows the rank order correlation plots 
    and the lower half shows a color coding of the Jaccard-index with warmer colors indicating higher index/correlation. 
    As can be seen, roughly half of the pairs show above-random correlation (grey color) 
    and two of them show high correlation.
    }
    \label{fig:scatterplot-matrix}
\end{figure}

Lastly, we explore the rank order correlation and the Jaccard index, in order to see if the representations have any correlation. We compared every representation to all others and present the results together with the Jaccard index in a scatterplot matrix which gives a visual overview of the amount of correlation, see \autoref{fig:scatterplot-matrix}. 

What stands out most in this matrix are the two pairs: CTM compared to TranFV and WVF compared to PVF. The high correlation between the respective pairs shows as a deep red color representing a high Jaccard-index and rank-number plots approaching the linear behaviour $y = x$. 
This result is not surprising, we already know that the TranFV and the CTM are similar, and the WFV is a one-dimensional part of the PFV. 

Aside from these prominent correlations, we are especially interested in correlations for the topological feature representation. 
Similar to what we had seen in the t-SNE plots, there is a correlation to the physical properties SFV and WFV even if it is not as strongly expressed. It seems like there is a stronger agreement for the very low and very highly ranked pairs, while the middle part is rather uncorrelated.

%-------------------------------------------------------------------------
\section{Conclusions}
In this case study, we applied some methods from Visual Analytics to analyze feature vector representations for ensemble data visualization in a chemistry application.
Our analysis gave us not only some insight into the correlation of the different representations in our specific case study but also, encourages generally to experiment more openly with novel feature descriptors and spend some time exploring their discriminative capabilities.
We didn't get final answers to our motivating questions about what constitutes a good feature representation but show possibilities towards this goal. 

With respect to our case study, some of the results were expected, for example, that the reduced charge transfer vector (TranFV) provides similar insights as the full charge transfer matrix (CTM). This is a valuable confirmation of the representation chosen in previous work~\cite{Thygesen2022}.
In contrast, the correlation of the topological representation with the physical properties was rather surprising. At first, it ignores the spatial segmentation that is the basis of all current analysis methods used by the domain scientists. Secondly, it only uses a simple combination of topological invariants.
This result is very encouraging to further explore topological methods in this application and experiment with more advanced topological feature descriptors. 
The performance of the scatterplot representation was overall rather poor, but also not entirely surprising since we used very coarse downsampling of the actual bi-variate distribution. We still think that the full distribution is very characteristic of the charge transfer, and plan to further investigate better ways to quantify it concisely. 

Beyond the specific application, the case study demonstrated the value of applying feature comparison methods, as typically used in Visual analytics applications.
We see a large potential in further exploiting the methods used in this study, and performing a more detailed inspection, for instance looking at the outliers in the t-SNE projections would be interesting.
Integrating the VA methods directly in an interactive exploration framework as presented in~\cite{Thygesen2022} with spatial representations of the scalar fields could further improve the understanding of the results obtained so far. 

\newpage
\balance

% ========================================================================================= %
% Acknowledgments
\acknowledgments{
We thank Mathieu Linares for providing the data used in the case study, and for the fruitful discussions. 
This work is supported by the SeRC (Swedish e-Science Research Center), the Swedish Research Council~(VR) grant 2019-05487, and the ELLIIT environment for strategic research in Sweden. S.S. Thygesen is associated with Wallenberg AI, Autonomous Systems and Software Program (WASP).

% From Level-of-Detail paper
%This work is supported by SeRC (Swedish e-Science Research Center), the Swedish Research Council~(VR) grant 2019-05487, and an Indo-Swedish joint network project: DST/INT/SWD/VR/P-02/2019 VR grant 2018-07085. VN is partially supported by a Swarnajayanti Fellowship from the Department of Science and Technology, India (DST/SJF/ETA-02/2015-16) and a Mindtree Chair research grant. 
%SST is associated to the Wallenberg AI, Autonomous Systems and Software Program (WASP).
%The computational resources were provided by the Swedish National Infrastructure for Computing (SNIC) at NSC (VR grant 2018-05973). 
}

\bibliographystyle{abbrv-doi}
%\bibliographystyle{abbrv-doi-narrow}
%\bibliographystyle{abbrv-doi-hyperref}
%\bibliographystyle{abbrv-doi-hyperref-narrow}

%\clearpage
\bibliography{bibliography}

\begin{thebibliography}{10}

\bibitem{AlSaadon2019}
R.~Al-Saadon, T.~Shiozaki, and G.~Knizia.
\newblock {Visualizing Complex-Valued Molecular Orbitals}.
\newblock {\em The Journal of Physical Chemistry A}, 123:3223--3228, 2019.

\bibitem{EMBCOMPARE4}
D.~L. Arendt, N.~Nur, Z.~Huang, G.~Fair, and W.~Dou.
\newblock {{P}arallel {E}mbeddings: A Visualization Technique for Contrasting
  Learned Representations}.
\newblock In {\em Proceedings of the 25th International Conference on
  Intelligent User Interfaces}, IUI~'20, pp. 259--274. ACM, 2020.

\bibitem{Chazal2021}
F.~Chazal and B.~Michel.
\newblock {An Introduction to Topological Data Analysis: Fundamental and
  Practical Aspects for Data Scientists}.
\newblock {\em Frontiers in Artificial Intelligence}, 2021.

\bibitem{edelsbrunner2008persistent}
H.~Edelsbrunner, J.~Harer, et~al.
\newblock Persistent homology-a survey.
\newblock {\em Contemporary mathematics}, 453:257--282, 2008.

\bibitem{Espadoto:21}
M.~Espadoto, R.~M. Martins, A.~Kerren, N.~S.~T. Hirata, and A.~C. Telea.
\newblock {Toward a Quantitative Survey of Dimension Reduction Techniques}.
\newblock {\em IEEE Transactions on Visualization and Computer Graphics},
  27(3):2153--2173, 2021.

\bibitem{Fieller1957}
E.~Fieller, H.~Hartley, and E.~Pearson.
\newblock Tests for rank correlation coefficients.
\newblock {\em Biometrika}, 44(3--4):470--481, 1957.

\bibitem{Haranczyk2008}
M.~Haranczyk and M.~Gutowski.
\newblock {Visualization of Molecular Orbitals and the Related Electron
  Densities}.
\newblock {\em Journal of Chemical Theory and Computation}, 4(5):689--693,
  2008.

\bibitem{EMBCOMPARE3}
F.~Heimerl, C.~Kralj, T.~Moller, and M.~Gleicher.
\newblock {{embComp}: Visual Interactive Comparison of Vector Embeddings}.
\newblock {\em IEEE Transactions on Visualization and Computer Graphics}, 2020.

\bibitem{Hotz2020}
I.~Hotz, R.~Bujack, C.~Garth, and B.~Wang.
\newblock {Mathematical Foundations in Visualization}.
\newblock In M.~Chen, H.~Hauser, P.~Rheingans, and G.~Scheuermann, eds., {\em
  Foundations of Data Visualization}. Springer, 2020.

\bibitem{JACCARD}
P.~Jaccard.
\newblock {The Distribution of the Flora in the Alpine Zone.1}.
\newblock {\em New Phytologist}, 11(2):37--50, 1912.

\bibitem{Katz2016}
G.~Katz, E.~C.~R. Shin, and D.~Song.
\newblock {ExploreKit: Automatic Feature Generation and Selection}.
\newblock In {\em IEEE 16th International Conference on Data Mining (ICDM)},
  2016.

\bibitem{VA1}
D.~A. Keim, J.~Kohlhammer, G.~Ellis, and F.~Mansmann, eds.
\newblock {\em Mastering the Information Age: Solving Problems with Visual
  Analytics}.
\newblock Eurographics Association, 2010.

\bibitem{KS:12}
A.~Kerren and F.~Schreiber.
\newblock {Toward the Role of Interaction in Visual Analytics}.
\newblock In {\em Proceedings of the Winter Simulation Conference}, WSC '12,
  pp. 420:1--420:13. Winter Simulation Conference, 2012.

\bibitem{EMBCOMPARE2}
S.~Liu, P.-T. Bremer, J.~J. Thiagarajan, V.~Srikumar, B.~Wang, Y.~Livnat, and
  V.~Pascucci.
\newblock {Visual Exploration of Semantic Relationships in Neural Word
  Embeddings}.
\newblock {\em IEEE Transactions on Visualization and Computer Graphics},
  24(1):553--562, Jan. 2018.

\bibitem{Martin2003NTO}
R.~L. Martin.
\newblock {Natural transition orbitals}.
\newblock {\em The Journal of Chemical Physics}, 118(11):4775--4777, 2003.

\bibitem{eurovis_electronic_densities}
T.~B. Masood, S.~S. Thygesen, M.~Linares, A.~I. Abrikosov, V.~Natarajan, and
  I.~Hotz.
\newblock {Visual Analysis of Electronic Densities and Transitions in
  Molecules}.
\newblock {\em Computer Graphics Forum}, 40(3):287--298, 2021.

\bibitem{norman2011perspective}
P.~Norman.
\newblock A perspective on nonresonant and resonant electronic response theory
  for time-dependent molecular properties.
\newblock {\em Physical chemistry chemical physics}, 13(46):20519--20535, 2011.

\bibitem{Post2003}
F.~H. Post.
\newblock {The State of the Art in Flow Visualization: Feature Extraction and
  Tracking}.
\newblock In {\em Computer Graphics Forum}, vol. 22(4), pp. 775--792. Blackwell
  Publishing Inc, 2003.

\bibitem{Sharma2021CSPpeeling}
M.~Sharma, T.~B. Masood, S.~S. Thygesen, M.~Linares, I.~Hotz, and V.~Natarajan.
\newblock {Segmentation Driven Peeling for Visual Analysis of Electronic
  Transitions}.
\newblock In {\em 2021 IEEE Visualization Conference (VIS)}, pp. 96--100, 2021.

\bibitem{EMBCOMPARE1}
D.~Smilkov, N.~Thorat, C.~Nicholson, E.~Reif, F.~B. Vi{\'e}gas, and
  M.~Wattenberg.
\newblock {{E}mbedding {P}rojector: Interactive Visualization and
  Interpretation of Embeddings}.
\newblock In {\em Proceedings of the NIPS 2016 Workshop on Interpretable
  Machine Learning for Complex Systems}, 2016.

\bibitem{Stone2011}
J.~E. Stone, D.~J. Hardy, J.~Saam, K.~L. Vandivort, and K.~Schulten.
\newblock {GPU-accelerated computation and interactive display of molecular
  orbitals}.
\newblock In {\em GPU Computing Gems Emerald Edition}, Applications of GPU
  Computing Series, pp. 5--18. Elsevier Inc., 2011.

\bibitem{Thygesen2022}
S.~S. Thygesen, T.~B. Masood, M.~Linares, V.~Natarajan, and I.~Hotz.
\newblock Level of detail exploration of electronic transition ensembles using
  hierarchical clustering.
\newblock {\em Computer Graphics Forum}, 41(3):333--344, 2022.

\bibitem{TSNE}
L.~van~der Maaten and G.~Hinton.
\newblock {Visualizing Data using t-SNE}.
\newblock {\em Journal of Machine Learning Research}, 9(86):2579--2605, 2008.

\bibitem{Wasserman2006}
L.~Wasserman.
\newblock {\em All of Nonparametric Statistics}.
\newblock Springer, 2006.

\bibitem{Whitaker2020}
R.~T. Whitaker and I.~Hotz.
\newblock {Transformations, Mappings, and Data Summaries}.
\newblock In M.~Chen, H.~Hauser, P.~Rheingans, and G.~Scheuermann, eds., {\em
  Foundations of Data Visualization}. Springer, 2020.

\bibitem{witschard2022}
D.~Witschard, I.~Jusufi, R.~M. Martins, K.~Kucher, and A.~Kerren.
\newblock Interactive optimization of embedding-based text similarity
  calculations.
\newblock {\em Information Visualization}, 21(4):335--353, 2022.

\bibitem{yan2021scalar}
L.~Yan, T.~B. Masood, R.~Sridharamurthy, F.~Rasheed, V.~Natarajan, I.~Hotz, and
  B.~Wang.
\newblock {Scalar field comparison with topological descriptors: Properties and
  applications for scientific visualization}.
\newblock In {\em Computer Graphics Forum}, vol.~40, pp. 599--633. Wiley Online
  Library, 2021.

\end{thebibliography}

\end{document}